\documentclass[letterpaper, 10 pt, conference]{ieeeconf}  

\IEEEoverridecommandlockouts                              
\usepackage[top=0.833in, bottom=0.597in, left=0.75in, right=0.75in]{geometry}                                

\usepackage{amsmath,amssymb,amsfonts}
\usepackage{algorithmic}
\usepackage{microtype}
\usepackage{graphicx}
\usepackage{textcomp}
\usepackage{xcolor}
\usepackage{booktabs}
\usepackage{cite}
\usepackage{hyperref}
\usepackage{bm}
\usepackage{mathtools}
\usepackage{algorithm}
\usepackage{algorithmic}

\newcommand{\R}{\mathbb{R}}

\newcommand{\calS}{\mathcal{S}}
\newcommand{\calC}{\mathcal{C}}        

\newcommand{\bfP}{\mathbf{P}}
\newcommand{\DKL}{D_{\mathrm{KL}}}
\newcommand{\kexh}{r_{\mathrm{exh}}}
\newcommand{\radh}{r_{\mathrm{adh}}}
\newcommand{\PS}[1]{\mathbf{P}_{S#1}}

\title{\LARGE \bf
Preparation of Papers for IEEE Sponsored Conferences \& Symposia*
}

\title{\LARGE \bf
Trajectory Landscapes for Therapeutic Strategy Design in Agent-Based Tumor Microenvironment Models
}

\author{Eric~Cramer,~Laura~M.~Heiser,~and~Young~Hwan~Chang%
  \thanks{
    E.~Cramer, L.~M.~Heiser, and Y.~H.~Chang are with the Department of Biomedical Engineering, Oregon Health \& Science University, Portland, OR 97239, USA. {\tt\small \{cramere, heiserl, chanyo\}@ohsu.edu}.
  }
}

\begin{document}

\maketitle
\thispagestyle{empty}
\pagestyle{empty}

\begin{abstract}
Multiplex tissue imaging~(MTI) enables high-dimensional, spatially resolved measurements of the tumor microenvironment~(TME), but most clinical datasets are temporally undersampled and longitudinally limited, restricting direct inference of underlying spatiotemporal dynamics and effective intervention timing. Agent-based models~(ABMs) provide mechanistic, stochastic simulators of TME evolution; yet their high-dimensional state space and uncertain parameterization make direct control design challenging. 
This work presents a reduced-order, simulation-driven framework for therapeutic strategy design using ABM-derived trajectory ensembles. Starting from a nominal ABM, we systematically perturb biologically plausible parameters to generate a set of simulated trajectories and construct a low-dimensional trajectory landscape describing TME evolution. From time series of spatial summary statistics extracted from the simulations, we learn a probabilistic Markov State Model~(MSM) that captures metastable states and the transitions between them. 
To connect simulation dynamics with clinical observations, we  map patient MTI snapshots onto the landscape and assess concordance with observed spatial phenotypes and clinical outcomes. We further show that conditioning the MSM on dominant governing parameters yields group-specific transition models to formulate a finite-horizon Markov Decision Process (MDP) for treatment scheduling. The resulting framework enables simulation-grounded therapeutic policy design for partially observed biological systems without requiring longitudinal patient measurements.
\end{abstract}


\section{INTRODUCTION}
\label{sec:intro}

The application of feedback and decision-making frameworks to biological systems has the potential to move oncology from fixed, open-loop treatment schedules toward adaptive, state-aware therapeutic strategies~\cite{sontag_new_2004, johnson_omic_2022}. In cancer, the tumor microenvironment (TME) evolves as a complex, spatially distributed stochastic dynamical system~\cite{iglesias_control_2009}, where tumor growth, immune infiltration, and stromal remodeling drive trajectories toward either immune control or immune-escape~\cite{dunn_three_2004}. 

Designing such strategies is challenging due to limited observability. Although multiplex tissue imaging (MTI) provides high-dimensional, spatially resolved measurements of the TME~\cite{keren_structured_2018}, clinical samples are sparse and provide irregular snapshots of an evolving process. As a result, most patient datasets are insufficient to infer the underlying spatiotemporal dynamics, identify transitions between latent dynamical states of the TME, or determine when therapeutic intervention is most effective.

Designing intervention strategies therefore requires models that relate therapeutic inputs to measurable physiological state. Ordinary differential equation (ODE) models remain widely used due to their analytical tractability, but their well-mixed assumption fails to capture the spatial heterogeneity revealed by MTI~\cite{sibai_spatial_2025}. Models that track only global population averages cannot represent spatially organized mechanisms critical to therapy response---such as T cell exclusion, compartment segregation, and localized exhaustion gradients. These limitations motivate approaches that explicitly capture spatiotemporal TME dynamics while enabling reduced-order abstractions suitable for decision-making.

Agent-based models (ABMs) provide a mechanistic framework for representing stochastic, spatially resolved cell–cell and cell–matrix interactions in the TME~\cite{ghaffarizadeh_physicell_2018,johnson_human_2025}. However, calibrating ABMs to clinical data remains difficult because temporally undersampled measurements of the TME provide only partial observations of the underlying dynamical state. Consequently, identifying a single optimal set of model parameters is often underdetermined and may fail to capture the range of dynamical behaviors relevant for therapeutic strategy design~\cite{cogno_agent-based_2024, wang_comprehensive_2024}. 

To address this limitation, we adopt a simulation-ensemble perspective in which we treat a nominal ABM as a generator of plausible system dynamics. By systematically perturbing key model parameters, we generate a large ensemble of simulated trajectories spanning biologically plausible regimes of TME evolution. This ensemble defines a trajectory landscape whose metastable regions serve as attractors of the TME dynamics, enabling the construction of reduced-order, discrete-state representations suitable for Markov modeling and control design. We validate this landscape by mapping clinical biospecimens to their nearest attractor states and associating the resulting assignments with clinical outcomes. A parameter sensitivity analysis then identifies which biological parameters govern attractor selection and partition the landscape into distinct dynamical regimes. Building on this reduced-order representation, we introduce a Markov State Model (MSM) to learn the transition structure among system states and formulate a finite-horizon Markov Decision Process (MDP) to optimize treatment schedules.

\section{Problem Formulation}
\label{sec:problem}

\vspace{-0.2cm}
\subsection{Agent-Based Model of the TME}
\label{sec:abm}
\begin{figure}[!t]
  \centering
  \includegraphics[width=0.75\columnwidth]{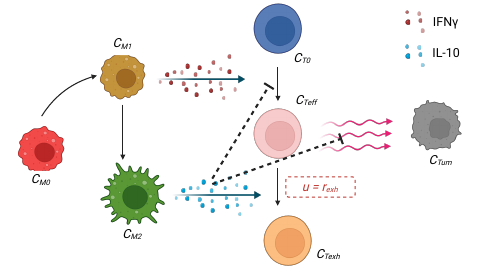}
  \vspace{-4mm}
  \caption{Agent-based TME signaling architecture. The model encodes seven cell types and the signaling interactions governing their transitions; the T-cell exhaustion rate $\kexh$ is the primary control-relevant parameter. T-cell exhaustion ($\kexh$) is a state transition rate governing the progressive loss of effector function in tumor-infiltrating T cells; immunotherapeutic interventions such as checkpoint blockade act in part by reducing this rate.}
  \vspace{-5mm}
  \label{fig:system}
\end{figure}

\vspace{-0.25cm}
We consider a discrete-time ABM of the TME, implemented in the
PhysiCell framework~\cite{ghaffarizadeh_physicell_2018} using the cell behavior
grammar of Johnson {\it et al.}~\cite{johnson_human_2025}. For simulation $i$ with parameter vector $\mathbf{p}^i$, the configuration of all agents defines the system state at time $k$,
\begin{equation}
\mathcal{A}_k^i = \{ a_{k,1}^i, a_{k,2}^i, \dots, a_{k,N_k^i}^i \},
\label{eq:abm_config}
\end{equation}
where $N_k^i$ is the time-varying number of agents. To define the observation map, we focus on the MTI-observable components of each agent---spatial location and cell type---and represent each agent as
\begin{equation}
a_{k,j}^i = \big(x_{k,j}^i,\, y_{k,j}^i,\, c_{k,j}^i\big)
\label{eq:agent_state_simple}
\end{equation}
where $(x_{k,j}^i,\, y_{k,j}^i) \in \R^2$ denotes the spatial
location of agent $j$, and $c_{k,j}^i \in \calC = \{ c_{\mathrm{tum}}, c_{\mathrm{T_0}}, c_{\mathrm{T_{eff}}}, c_{\mathrm{T_{exh}}}, c_{\mathrm{M0}}, c_{\mathrm{M1}}, c_{\mathrm{M2}} \}$ denotes its cell type. The ABM evolves according to a stochastic update rule governed by local behavioral interactions:
\begin{equation}
\mathcal{A}_{k+1}^i = \mathcal{F}\!\left(\mathcal{A}_k^i,\, \mathbf{p}^i,\, \xi_k \right),
\label{eq:abm_dynamics}
\end{equation}
where $\mathcal{F}$ denotes the simulator update map and $\xi_k$ represents stochastic draws associated with the behavioral rules governing proliferation, apoptosis, migration, and phenotypic transitions. The interaction rules encoded in $\mathcal{F}$ implement the signaling architecture illustrated in Fig.~\ref{fig:system}. Because these rules act on individual agents, the resulting system state is high-dimensional and varies over time.
A key open question is which of the sampled parameters in $\mathbf{p}^i$ most strongly determine the long-run behavior (i.e., the terminal attractor state reached after sufficient simulation time) of the system.



\subsection{Observation Map via Spatial Statistics}
To enable comparison between simulations and MTI data, and to account for the variable number of cells in each configuration, we represent each agent configuration $\mathcal{A}_k^i$ (the state of the ABM at time step $k$) using a fixed-dimensional vector of spatial summary statistics and composition features. Specifically, we define an observation map that transforms each configuration of agents into a feature vector:
\begin{equation}
u_k^i = \mathcal{U}\!\left(\mathcal{A}_k^i\right) \in \mathbb{R}^{M},
\label{eq:obs_map}
\end{equation}
where $M$ denotes the number of spatial statistics extracted by $\mathcal{U}(\cdot)$, $i = 1, \ldots, N$ indexes simulations, and $k = 1, \ldots, K$ indexes time steps. The $\ell$-th component $u_{k,\ell}^i$ denotes a single spatial statistic, for $\ell = 1, \ldots, M$. These features include cell-type proportions, spatial mixing or segregation metrics, neighborhood interaction statistics, and graph-based descriptors of tissue organization.
Stacking these observations across time yields a sequence
\begin{equation}
U^i = \left[ u_1^i, u_2^i, \dots, u_K^i \right] \in \mathbb{R}^{M \times K}.
\label{eq:obs_stack}
\end{equation}
which provides a fixed-dimensional representation of each simulation trajectory $\mathbf{p}^i$ suitable for clustering and Markov State Model (MSM) estimation.



\subsection{Temporal Embedding to Capture Dynamics}
\label{sec:embed}
Because $u_k^i$ is a partial observation of the ABM state, treating each time point independently may lead to  clustering that ignores temporal continuity within a trajectory. To incorporate temporal context, we adopt a delay-embedding approach inspired by Takens' theorem~\cite{rand_detecting_1981}. Although Takens' theorem formally applies to smooth deterministic systems, delay embedding is used here as a practical reconstruction heuristic for recovering latent dynamics from partial observations. Prior work has demonstrated the effectiveness of this approach in discrete biological systems~\cite{copperman_morphodynamical_2023}, and we empirically validate its applicability to our stochastic ABM in Section~\ref{sec:results}. We construct a local time-window embedding by stacking observations within a symmetric window of radius $w$:
\begin{equation}
\mathcal{W}_k^i =
\left[
\left(u_{k-w}^i\right)^\top,
...
\left(u_{k}^i\right)^\top,
...
\left(u_{k+w}^i\right)^\top
\right]^\top
\in \mathbb{R}^{M(2w+1)}.
\label{eq:time_embedding}
\end{equation}
$\mathcal{W}_k^i$ captures short-horizon temporal variation in spatial statistics, providing a richer representation of the system's local dynamical behavior than a single observation alone.

We pool the embedded vectors $\{\mathcal{W}_k^i\}$ across simulations (generated from parameter perturbations) and time, and cluster them to obtain a discrete set of TME states. Since each $\mathcal{W}_k^i$ represents a local temporal window of spatial statistics rather than a single snapshot, the resulting clusters correspond to characteristic dynamical regimes (metastable states) rather than static tissue configurations. The clustered embedding therefore defines a trajectory landscape of TME evolution that can be used to estimate transition dynamics and analyze intervention strategies.

\begin{figure}[!t]
  \centering
  \includegraphics[width=0.75\columnwidth]{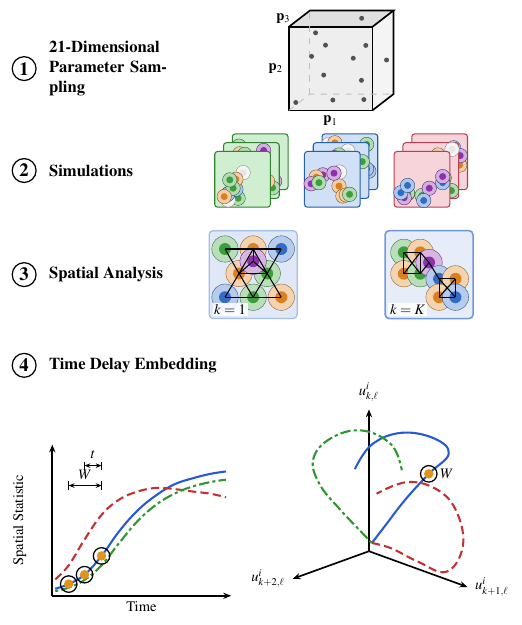}
  \vspace{-4mm}
  \caption{Simulation-to-embedding pipeline. (1)~Latin hypercube sampling of the parameter space yields $N$ parameter vectors $\mathbf{p}^i$. (2)~Each vector seeds a PhysiCell simulation producing a cell-configuration trajectory. (3)~Spatial statistics $u_k^i \in \mathbb{R}^M$ are extracted at each step $k = 1,\ldots,K$. (4)~Delay-coordinate embedding stacks a window of $w$ consecutive observations, reconstructing latent dynamical structure from the scalar spatial-statistic time series.}
  \vspace{-5mm}
  \label{fig:overview}
\end{figure}

\section{Method}
\label{sec:methodology}

\subsection{Simulation Ensemble Construction}
\label{sec:params}
\vspace{-1mm}
To explore the range of possible tumor microenvironment dynamics, we generate a simulation ensemble by sampling the ABM parameter space using a Latin Hypercube design. We sample 50 distinct parameter combinations across a 21-dimensional parameter space spanning cell motility rates, adhesion strengths, and signaling molecule secretion parameters (Fig.~\ref{fig:overview}.1).
We simulated three randomized spatial initializations of cell populations for each parameter set, yielding $3N$ total trajectories that capture a diverse set of biologically plausible dynamical regimes (Fig.~\ref{fig:overview}.2).


\subsection{Time Delay Embedding and State Discretization}
\label{sec:tdembed}
\vspace{-0.1cm}
For each simulation trajectory, we construct delay-embedded feature vectors using the temporal window representation described in Section \ref{sec:problem} (Fig.~\ref{fig:overview}.3-4). Specifically, the embedded vector $\mathcal{W}_k^i$ stacks $W{=}50$ consecutive observations. To visualize the trajectory structure, we project the embedded vectors onto a 2D UMAP manifold (Fig.~\ref{fig:landscape})~\cite{mcinnes_umap_2020}. Hierarchical clustering identifies metastable TME states. This procedure yields a discrete set of metastable TME states $\calS = \{S_1, \ldots, S_6\}$ representing characteristic dynamical regimes of tumor evolution. 

\subsection{Parameter Identification}
\label{sec:param_id_method}
\vspace{-0.10cm}
To determine which model parameters govern long-run trajectory behavior, we perform a Kruskal--Wallis nonparametric test for each sampled parameter. The test compares the distribution of parameter values across simulations grouped by their final cluster assignment (Section~\ref{sec:msm}).

We identify parameters with statistically significant separation between groups as dominant drivers of attractor selection. We subsequently use these parameters to define group-conditioned transition models in the Markov state analysis (Section~\ref{sec:msm_results}).
\subsection{Markov State Model Estimation}
\label{sec:msm}
\vspace{-0.10cm}
We model the discrete state sequences obtained from trajectory clustering as a first-order Markov chain, estimating transition probabilities from observed transition counts $\hat{P}_{ij} = N_{ij} / \sum_l N_{il}$ where $N_{ij}$ denotes the number of observed transitions from state $S_i$ to state $S_j$.
For each attractor group ($S_4$-prone or $S_6$-prone), we estimate separate transition matrices $\hat{\PS{g}}$, where $g \in \{4, 6\}$ indexes the terminal attractor state, using trajectories stratified by the identified parameter regimes. In addition, we estimate a shared drug-condition transition matrix $\hat{\PS{1}}$ from simulations with reduced exhaustion rate $\kexh$, representing the effect of Drug A.
\subsection{MDP Formulation for Intervention Design}
\label{sec:mdp}
\vspace{-0.10cm}
The learned transition models define a finite-horizon MDP over horizon $K$ with state space $\calS$ and binary action set $a \in \{0\text{ (no drug)},\, 1\text{ (Drug A)}\}$. The action-conditioned transition matrices are $\bfP^{(0)} = \PS{g(s)}$ and $\bfP^{(1)} = \PS{1}$ (Section~\ref{sec:msm}), where $g(s) \in \{S_4\text{-prone},\, S_6\text{-prone}\}$ denotes the attractor group associated with state $s$. The control objective is to maximize absorption into $S_1$ via the terminal reward $\rho(S_1) = 1$, $\rho(S_j) = 0$ for $j \neq 1$.

The optimal value function $V(s, k)$ satisfies Bellman's equation
\begin{equation}
  V(s, k) = \max_{a \in \{0,1\}} \sum_{s' \in \calS} [\bfP^{(a)}]_{ss'}\, V(s', k{+}1),
  \label{eq:bellman}
\end{equation}
which we solve via backward induction from the terminal condition $V(s, K) = \rho(s)$. 
The resulting policy  $\mu^*(s,k) = \arg\max_{a}\sum_{s'} [\bfP^{(a)}]_{ss'}\, V(s', k{+}1)$ applies Drug~A only when doing so increases expected probability of eventual absorption into the effector T cell-dominant state ($S_1$).

\section{Results}
\label{sec:results}

\subsection{Ensemble Trajectory Landscape}
\begin{figure}[hbt!]
  \centering
  \includegraphics[width=\columnwidth]{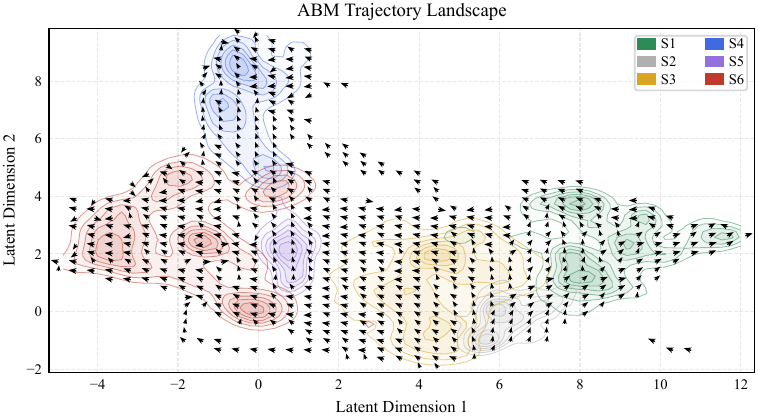}
  \vspace{-7mm}
  \caption{ABM Simulation Trajectory Landscape. The vector field indicates the mean intra-window displacement in UMAP space. Contours show the density of each metastable TME state ($S_1$--$S_6$). Trajectories diverge from a shared early origin toward terminal attractors.}
  \vspace{-3mm}
  \label{fig:landscape}
\end{figure}

\vspace{-0.1cm}
Visualization of the delay-embedded landscape reveals clear structure in the simulated trajectories. Simulations originate from a shared region in the latent space and diverge toward six stable clusters (Fig.~\ref{fig:landscape}). The vector field indicates the mean local displacement of trajectories in the embedding, while contour lines represent the density of observed TME states. 



\begin{figure}[hbt!]
  \centering
  \includegraphics[width=0.75\columnwidth]{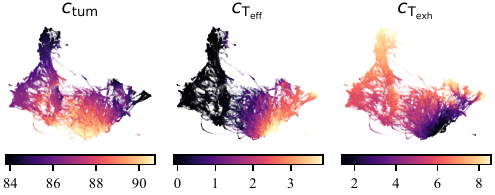}
  \vspace{-3mm}
  \caption{UMAP landscape colored by relative cell-type population levels: (A) $c_{\mathrm{tum}}$, (B) $c_{\mathrm{T_{eff}}}$, (C) $c_{\mathrm{T_{exh}}}$. Each point is a delay-embedded simulation window. $c_{\mathrm{tum}}$ is elevated in the immune-escape region, $c_{\mathrm{T_{eff}}}$ concentrates near $S_1$, and $c_{\mathrm{T_{exh}}}$ accumulates toward $S_4$ and $S_6$.}
  \vspace{-5mm}
  \label{fig:pop_features}
\end{figure}

Three terminal attractors emerge: $S_1$, a state of effector T cell dominance; $S_4$,a state of exhausted T cell dominance; and $S_6$, a state of T cell exclusion from the tumor (Fig.~\ref{fig:pop_features}). The remaining states $S_2$, $S_3$, and $S_5$ serve as transient waypoints capturing immune evolution from the initial region towards dysfunctional terminal regimes. 

\subsection{Synthetic Mapping and Clinical Validation}
\vspace{-2mm}

We evaluated whether latent TME states can be inferred from a single static observation with a synthetic mapping experiment. Specifically, we generated test datasets by randomly sampling 15\% of ABM time points from each trajectory and repeating this procedure 1000 times. 

To align the observation space with the clinical MTI data (in which macrophage functional status markers were not measured), we removed macrophage-associated features from the observation vectors for this analysis. We then assigned each sampled snapshot to the nearest delay-embedded state using 1-nearest-neighbor classification with cosine similarity. 
One-vs-rest ROC analysis yielded per-state AUC values of 0.976 ($S_1$), 0.950 ($S_2$), 0.845 ($S_3$), 0.982 ($S_4$), 0.981 ($S_5$), and 0.978 ($S_6$), with a mean AUC of 0.952, demonstrating that the mapping from snapshot-level spatial statistics to the learned dynamical state representation remains discriminative even when macrophage features are excluded. These results indicate that the learned embedding generalizes to the macrophage-excluded feature set available in clinical data.

\begin{figure}[!t]
  \centering
  \includegraphics[width=\columnwidth]{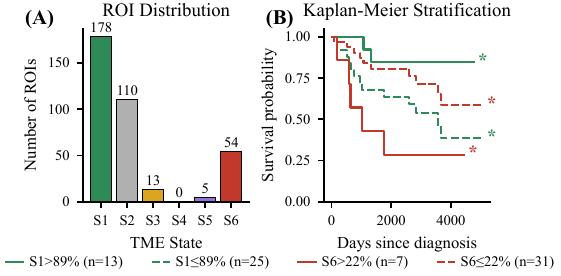}
  \vspace{-7mm}
  \caption{Clinical validation in the MIBI cohort (Keren et al., 2018; $n=38$ patients, 360 ROIs). (A) Distribution of MIBI ROIs across TME states $S_1$--$S_6$; $S_4$ receives zero clinical ROIs. (B) Kaplan-Meier survival curves stratified by log-rank-optimized $S_1$ and $S_6$ proportion cutoffs.}
  \vspace{-5mm}
  \label{fig:clinical}
\end{figure}

We next applied the learned state mapping to a data set of Multiplexed Ion Beam Imaging (MIBI) of clinical biospecimens collected from patients undergoing surgery for Triple Negative Breast Cancer~\cite{keren_structured_2018}. We projected 360 regions of interest (ROIs) from 38 patients onto the learned trajectory landscape. The majority of ROIs were assigned to either the effector T cell-dominant state $S_1$ ($n=178$) or the immune excluded state $S_6$ ($n=54$) as shown in Fig.~\ref{fig:clinical}. The exhausted T cell-dominant state $S_4$, observed in simulation, was not detected in this cohort. 

To assess prognostic value and clinical relevance, we examined the relationship between state occupancy and patient survival. Kaplan-Meier analysis using exploratory log-rank-optimized thresholds with the lifelines KaplanMeierFitter algorithm~\cite{davidson-pilon_lifelines_2019} shows that patients with a high fraction of $S_1$ ROIs ($>89\%$, $n=13$) exhibit significantly improved overall survival ($p=0.038$), whereas patients with a high fraction of $S_6$ ROIs ($>22\%$, $n=7$) exhibit significantly worse survival ($p=0.029$, Fig.~\ref{fig:clinical}B). These findings suggest that the inferred dynamical states capture clinically meaningful TME regimes and provide evidence that the simulation-derived landscape reflects biologically-relevant system behavior.

\subsection{Parameter Identification and Attractor Basins}
\label{sec:param_id}
\vspace{-0.1cm}
We determined the model parameters that govern desired trajectory outcomes with Kruskal--Wallis tests across all 21 sampled parameters, comparing parameter distributions between simulations grouped by their terminal cluster assignment ($n = 150$; $S_1$: $n{=}11$, $S_4$: $n{=}66$, $S_6$: $n{=}73$). Two parameters show strong statistical separation: the T-cell exhaustion rate $\kexh$ ($H = 31.5$, $p = 1.4{\times}10^{-7}$) and tumor self-adhesion $\radh$ ($H = 110.0$, $p = 1.3{\times}10^{-24}$). The remaining 19 parameters exhibit weak separation ($H < 11$, $p > 0.004$) and do not significantly influence terminal state selection.

The two parameters play distinct roles in determining system behavior. The exhaustion rate $\kexh$ primarily determines whether immune escape occurs: low values of $\kexh$ produce  T cell-dominant trajectories that converge to $S_1$, while higher values drive trajectories toward escape regimes. Conditioned on high $\kexh$, the tumor self-adhesion parameter $\radh$ determines which escape mode emerges. High adhesion produces cohesive tumor clusters that resist immune infiltration, leading to the exhausted T cell-dominant attractor ($S_4$). In contrast, low adhesion results in diffuse tumor configurations associated with T cell exclusion and convergence to $S_6$ in which T cells may penetrate too quickly, leading to loss of function-supporting niches and dying out~\cite{jenkins_current_2023}.

\begin{figure}[!t]
  \centering
  \includegraphics[width=0.85\columnwidth]{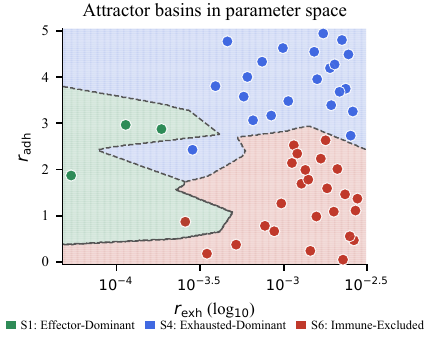}
  \caption{Attractor basin map in $(\log_{10}\kexh,\, \radh)$ parameter space. Shaded regions show the 2-NN decision boundary (99.3\% LOO-CV accuracy, $n{=}150$).}
  \vspace{-8mm}
  \label{fig:basins}
\end{figure}

To quantify this structure, we train a 2-nearest-neighborclassifier using leave-one-out cross-validation in the two-dimensional parameter space ($\log_{10}\kexh,\, \radh)$. The classifier achieves $99.3\%$ accuracy, indicating that the parameter space is effectively partitioned into three contiguous attractor basins (Fig.~\ref{fig:basins}). This strong separability suggests that identifying a patient's parameter regime---using measurable proxies such as exhaustion marker expression for $\kexh$ and spatial compactness metrics for $\radh$---may be sufficient to predict which escape mode is most relevant for therapeutic intervention. 

\subsection{MSM Construction and Validation}
\label{sec:msm_results}
\vspace{-0.15cm}
Building on the parameter axes identified in the previous analysis, we construct Markov State Model (MSM) conditioned on the terminal attractor group. This yields three separate $6{\times}6$ transition matrices:
$\PS{1}$ (effector T cell-dominant, $n{=}11$ simulations),
$\PS{4}$ (exhausted T cell-dominant, $n{=}66$), and
$\PS{6}$ (T cell-excluded, $n{=}73$).
Conditioning the MSM on attractor groups captures the distinct dynamical pathways through the transient landscape that lead to each escape mode, which a pooled transition model would obscure.

\textbf{Escape-mode matrices.}
For trajectories in the S4-prone group, escape proceeds through an exhaustion-driven pathway, reflected in the transition matrix:
\[
\hat{\PS{4}} = \tiny{ \begin{bmatrix}
0.997 & 0.000 & 0.000 & 0.003 & 0.000 & 0.000 \\
0.005 & 0.987 & 0.008 & 0.000 & 0.000 & 0.000 \\
0.001 & 0.001 & 0.994 & 0.001 & 0.003 & 0.001 \\
0.000 & 0.000 & 0.000 & 1.000 & 0.000 & 0.000 \\
0.000 & 0.000 & 0.000 & 0.004 & 0.994 & 0.002 \\
0.000 & 0.000 & 0.000 & 0.006 & 0.001 & 0.993
\end{bmatrix}}
\]
$S_4$ is the unique absorbing state, with the strongest inflow from the pre-exhaustion state $S_5$ ($p_{54}{=}0.004$) and the immune-excluded state $S_6$ ($p_{64}{=}0.006$), reflecting a drift toward exhausted T cell dominance as effector T cells progressively lose proliferative capacity and the effector population collapses.
In contrast, trajectories in the S6-prone group follow a spatial T cell exclusion pathway characterized by the transition matrix: 
\[
\hat{\PS{6}} = \tiny{ \begin{bmatrix}
0.997 & 0.000 & 0.000 & 0.000 & 0.000 & 0.003 \\
0.001 & 0.989 & 0.010 & 0.000 & 0.000 & 0.000 \\
0.000 & 0.000 & 0.995 & 0.000 & 0.001 & 0.004 \\
0.000 & 0.000 & 0.000 & 0.992 & 0.000 & 0.008 \\
0.000 & 0.000 & 0.000 & 0.001 & 0.993 & 0.007 \\
0.000 & 0.000 & 0.000 & 0.000 & 0.000 & 1.000
\end{bmatrix}}
\]
$S_6$ is the unique absorbing state; the largest probability fluxes toward it originate from the nominally distinct escape attractor $S_4$ ($p_{46}{=}0.008$) and the pre-exhaustion state $S_5$ ($p_{56}{=}0.007$), indicating that a number of simulations entering the exhaustion regime are ultimately redirected toward immune exclusion as the T cell population itself collapses. This aligns with prior literature finding T cell exhaustion often precedes immune exclusion from solid tumors~\cite{jenkins_current_2023}.
The structural differences between these matrices indicate that $S_4$ and $S_6$ represent fundamentally different escape regimes. Each regime is associated with distinct transient-state pathways, implying that effective interventions must target the specific dynamical route leading to escape.

\textbf{Drug-condition matrix.}
To model the effect of therapy, we construct a drug-conditioned transition matrix by reducing the T-cell exhaustion rate $\kexh$, which serves as a proxy for exhaustion-targeting immunotherapies. Under this condition (Drug~A), the reachable state space collapses to $\{S_1, S_2\}$: $S_1$ becomes fully absorbing, while $S_2$ gradually transition into $S_1$. The remaining states $S_3$--$S_6$ are not visited and therefore exhibit trivial self-transitions:
\[
\hat{\PS{1}} = \tiny{ \begin{bmatrix}
1.000 & 0.000 & 0.000 & 0.000 & 0.000 & 0.000 \\
0.007 & 0.993 & 0.000 & 0.000 & 0.000 & 0.000 \\
0.000 & 0.000 & 1.000 & 0.000 & 0.000 & 0.000 \\
0.000 & 0.000 & 0.000 & 1.000 & 0.000 & 0.000 \\
0.000 & 0.000 & 0.000 & 0.000 & 1.000 & 0.000 \\
0.000 & 0.000 & 0.000 & 0.000 & 0.000 & 1.000
\end{bmatrix}}
\]
$S_1$ is the unique absorbing state, receiving probability flux exclusively from $S_2$ ($p_{21}{=}0.007$); states $S_3$--$S_6$ exhibit trivial self-loops because they are unreachable under the reduced exhaustion rate, reflecting the collapse of the reachable state space to $\{S_1, S_2\}$.
This shift reflects the therapeutic mechanism whereby reducing exhaustion redirects system trajectories toward the effector T cell-dominant basin.

\textbf{Model validation.}
Bootstrap resampling (2000 resamples, $n{=}11$) yields all 95\% CI widths below 0.10 for entries of $\PS{1}$, confirming stable estimation. The Kullback-Leibler divergence ($\DKL$) between ABM-empirical and MSM-predicted occupancies remains bounded ($\DKL < 0.16$) across the simulation horizon. In addition, Chapman--Kolmogorov consistency tests show that the difference between the predicted and directly estimated multistep transition probabilities satisfies 
\begin{equation} 
\max_{i,j}|[\hat{\bfP}^n]_{ij} -
[\hat{\bfP}^{(n)}]_{ij}| < 1.2{\times}10^{-3}
\end{equation}
for $n{\in}\{2,5\}$ where $\hat{\mathbf{P}}^n$ is the $n$-th power of the one-step estimate and $\hat{\mathbf{P}}^{(n)}$ is estimated directly from $n$-step transitions. Together, these results support the validity of the first-order Markov approximation for the learned state representation.

\subsection{Intervention Analysis}
\label{sec:intervention}
\vspace{-1mm}

\textbf{Group-specific committor functions and time-sensitivity.}
Conditioning the MSM on escape mode enables analysis of group-specific intervention timing.
Let $\mathbf{q} \in [0,1]^6$ denote the vector of absorption probabilities into the effector T cell dominant state $S_1$ under the drug-conditioned transition matrix $\hat{\PS{1}}$, where $q_i = P(\text{reach }S_1 \mid X_0{=}S_i;\,\hat{\PS{1}})$.
Under the drug-conditioned dynamics, state $S_1$ is fully absorbing and $S_2$ transitions deterministically toward $S_1$, while $S_3$--$S_6$ are not visited and therefore exhibit trivial self-transitions. Consequently, the absorption probability vector becomes $\mathbf{q} = [1,1,0,0,0,0]^\top$.
To evaluate the effect of treatment timing, we define the group-specific committor function 
\begin{equation}
Q_{g}(k) = \boldsymbol{\pi}_{g}(k)\, \mathbf{q}
\end{equation}
where $\boldsymbol{\pi}_g(k) = \boldsymbol{\pi}_0^{(g)}\,(\PS{g})^k$ denotes the state distribution at step $k$ under the untreated group dynamics for group $g$. The quantity $Q_g(k)$ therefore represents the probability of eventual absorption into $S_1$ if Drug~A is administered at time step $k$.

\begin{figure}[!t]
  \centering
  \includegraphics[width=\columnwidth]{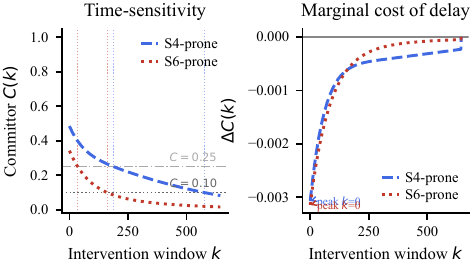}
  \vspace{-7mm}
  \caption{Intervention time-sensitivity by escape-mode group.
    \textbf{(left)} Committor curves $Q_g(k)$ for S4-prone and S6-prone groups; dashed lines mark the 25\% and 10\% thresholds.
    \textbf{(right)} Derivative $\mathrm{d}Q_g/\mathrm{d}k$: marginal cost of delaying intervention by one window.}
    \vspace{-5mm}
  \label{fig:timesens}
\end{figure}

The two escape-mode groups exhibit markedly different intervention sensitivities (Fig.~\ref{fig:timesens}). For the $S_4$-prone group, the initial intervention probability is $Q_g(0)$ = 0.485, while for the $S_6$-prone group it is $Q_g(0)=0.343$. However, the therapeutic window closes substantially faster in the $S_6$-prone regime. The probability of successful redirection falls below 25\% at $k=36$ for $S_6$-prone trajectories, compared with $k=189$ for the $S_4$-prone group, indicating that the T cell-exclusion pathway progresses much more rapidly toward an irreversible attractor.

\begin{center}
\footnotesize
\begin{tabular}{@{}lccc@{}}
  \toprule
  Group & $Q_g(0)$ & $Q_g(k){<}0.25$ at & $Q_g(k){<}0.10$ at \\
  \midrule
  S4-prone & 0.485 & $k = 189$ & $k = 577$ \\
  S6-prone & 0.343 & $k = 36$  & $k = 162$ \\
  \bottomrule
\end{tabular}
\end{center}

This difference reflects the faster concentration of probability mass in the T cell-exclusion basin under $\PS{6}$. Consistent with this observation, the derivative $\mathrm{d}Q_g/\mathrm{d}k$ reaches its largest magnitude at $k{=}0$ for both groups, indicating that the marginal cost of delaying treatment is greatest at the earliest observable time.

\textbf{MDP-optimal policy and strategy comparison.}
Four strategies are compared (Table~\ref{tab:strategies}): (1) no treatment, (2) an alternating drug-holiday schedule with cycle length $L{=}50$ windows, (3) an immediate full-dose switch at $k{=}0$, and (4) the optimal
state-dependent MDP policy $\mu^*$.
\setlength{\parskip}{0pt}

\begin{table}[hbt!]
  \caption{Strategy Comparison: Expected $P(\text{reach }S_1)$}
  \label{tab:strategies}
  \centering
  \renewcommand{\arraystretch}{1.0}
  \footnotesize
  \begin{tabular}{@{}lcc@{}}
    \toprule
    Strategy & S4-prone & S6-prone \\
    \midrule
    No treatment          & 0.083 & 0.017 \\
    Alternating ($L{=}50$)  & 0.194 & 0.079 \\
    Immediate ($k{=}0$)   & 0.485 & 0.343 \\
    \textbf{Optimal MDP}  & \textbf{0.632} & \textbf{0.401} \\
    \bottomrule
  \end{tabular}
\end{table}


Across both escape-modes, the optimal MDP policy achieves the highest probability of redirection to the effector T cell-dominant state $S_1$. Compared with the best fixed-timing strategy (immediate
treatment), the optimal MDP policy gains $+14.7\%$ (S4-prone) and $+5.8\%$ (S6-prone) in absorption probability. The policy applies Drug A only when it increases $V(s,k)$, withholding treatment in states already progressing toward $S_1$; fixed schedules cannot exploit this state-dependent structure.
\setlength{\parskip}{0pt}

\textbf{Simulation validation.}
To validate the MDP policy under closed-loop dynamics, we apply each treatment strategy to all simulated trajectories. For each trajectory, we propagate the initial state $s_0$ for $K = 646$ windows (matching the full trajectory length) using the transition matrices specified by the policy, and aggregate the resulting state distributions across simulations. The trajectory-level replay closely matches the theoretical predictions obtained from backward induction. Under the optimal MDP policy, 63.2\% of $S_4$-prone trajectories and 40.1\% of $S_6$-prone trajectories are redirected to the effector T cell-dominant state $S_1$. In absolute terms, this corresponds to 42 of 66 $S_4$-prone simulations and 29 of 73 $S_6$-prone simulations transitioning to immune control ($S_1$), compared with zero such transitions under the untreated dynamics. These results confirm that the reduced-order MSM provides an accurate stochastic model for evaluating intervention policies and that state-dependent control substantially improves therapeutic outcomes relative to fixed treatment schedules.


\section{Conclusion}
\label{sec:conclusion}
We convert a mechanistic ABM into a tractable switched Markov model for therapeutic strategy design. Delay embedding enables state assignment from partial spatial observations, and mapping clinical tumor samples onto the learned landscape shows that the inferred states are consistent with observed tumor phenotypes and associated with patient outcomes. Among 21 sampled parameters, only two govern long-term behavior: the T-cell exhaustion rate $\kexh$ and tumor self-adhesion $\radh$. Finite-horizon MDP optimization then yields state-dependent policies that redirect $42$ of $66$ S4-prone and $29$ of $73$ S6-prone simulations to immune control. The faster loss of therapeutic opportunity in the $S_6$-prone group further shows that early identification of the escape regime is critical.


\section*{Acknowledgements}
\small{
\label{sec:Acknowledgements}
This work was supported in part by NIH/NCI 1T32CA254888, NIH/NIGMS 5T32GM141938, NIH/ORIP S10OD034224, and the Jayne Koskinas Ted Giovanis Foundation for Health and Policy.
}


\bibliographystyle{IEEEtran}
\bibliography{my_library}

\end{document}